# Magnetic properties of Ge, Re and Cr substituted $Fe_5SiB_2$

M. Casadei[a,b], M.M. Isah[a], R. Cabassi[b], G. Trevisi[b], S. Fabbrici[b], M. Belli[b], C. de Julián Fernández[b], G. Allodi[c], V. Fournée[d], S. Sanna[a,*], F. Albertini[b].

[a] Department of Physics, University of Bologna, viale Berti Pichat 6/2, Bologna, Italy
[b] IMEM-CNR, Via delle Scienze 37, Parma, Italy
[c] Department of Mathematical, Physical and Computer Science, University of Parma, Parco Area delle Scienze 7/A, Parma, Italy
[d] Institut Jean Lamour, CNRS-Université de Lorraine, Campus Artem, 2 allée André Guinier, 54011 Nancy, France

*Corresponding author. Department of Physics, University of Bologna, viale Berti Pichat 6/2, Bologna, Italy.
E-mail address: s.sanna@unibo.it

## Abstract

One of the possible approaches to decrease the demand for critical elements such as rare earths is to develop new sustainable magnets. Iron-based materials are suitable for gap magnets applications since iron is the most abundant ferromagnetic element on Earth. $Fe_5SiB_2$ is a candidate as gap magnet thanks to its high Curie temperature ($T_C$ ~ 800 K) and saturation magnetization ($M_S$ ~ 140 $Am^2kg^{-1}$). However its anisotropy field is too low for applications ($H_A$ ~ 0.8 T). In order to increase the anisotropy value, we synthesized a series of Ge, Re and Cr substituted $Fe_5SiB_2$ samples and studied their magnetic properties. They all crystallize in the $Cr_5B_3$-type tetragonal structure with the I4/mcm space group. Curie temperature ($T_C$ = 803 K) and saturation magnetization ($M_S$ = 138 $Am^2kg^{-1}$) are slightly decreased by elemental substitution with Re having the largest effect. Despite being reduced, $T_C$ and $M_S$ still maintain significant values ($T_C$ > 750 K and $M_S$ = 118 $Am^2kg^{-1}$). The room temperature anisotropy field has been measured by Singular Point Detection (SPD) and increases by about 15% upon Re substitution, reaching 0.92 T for $Fe_{4.75}Re_{0.25}SiB_2$. We have also used Nuclear Magnetic Resonance and SPD measurements to study the spin reorientation transition which takes place at 172 K and we have found that it is partially suppressed by substitution of Ge from 172 K to 140 K and completely suppressed upon Cr and Re substitution.

## 1. Introduction

Permanent magnets play a key role in the transition from fossil fuels towards renewable energy alternatives, since they are largely used in electric motors, generators and energy converters [1], [2]. The permanent magnet market, at the current state, is essentially split into two types of materials. On one end, we have cheap but low performing ferrite magnets, with energy product lower than 50 $kJ/m^3$, $\mu_0 M_S \approx 0.45$ T and magnetocrystalline anisotropy energy MAE ≈ 0.33 $MJ/m^3$. On the other end, we have the more expensive and better performing rare earth (RE) magnets. RE magnets have unparalleled magnetic properties, due to the high magnetization of 3d electrons of transition metals ($\mu_0 M_S$ = 1.6 T) and the high anisotropy value granted by the enhanced spin-orbit interaction of 4f electrons of lanthanides (MAE = 4.9 $MJ/m^3$) [3]. This allows RE magnets to have energy products approaching 500 $kJ/m^3$ [4]. However, in the last decade, more concerns have been raised regarding the supply risks, volatile prices and the environmental impact of RE elements extraction and processing. In order to reduce the demand for critical materials, it is of fundamental importance to develop a new class of cheap magnets, called "gap magnets" [5], possessing intermediate properties, that could replace RE magnets in those applications that do not require very high performances. Since iron is one of the most abundant elements on Earth, iron-based magnets are desirable for this new class of materials. Fe$_5$SiB$_2$ is a notable candidate as a gap magnet since it possesses high saturation magnetization ($M_S \approx 140$ Am$^2$kg$^{-1}$) and Curie temperature ($T_C \approx 790$ K) [6]. It crystallizes in the body centered tetragonal (bct) Cr$_5$B$_3$-type structure with space group I4/mcm as shown in Figure 1 [7]. It presents uniaxial magnetocrystalline anisotropy with easy magnetization direction along the "c" axis at room temperature [8], which is of fundamental importance for permanent magnets, and undergoes a spin reorientation transition, with spins oriented on the basal plane for temperatures lower than $T_{SR} = 172$ K [9]. However, its magnetic anisotropy energy (MAE $\approx 0.3$ MJ/m$^3$) is too low for practical applications.

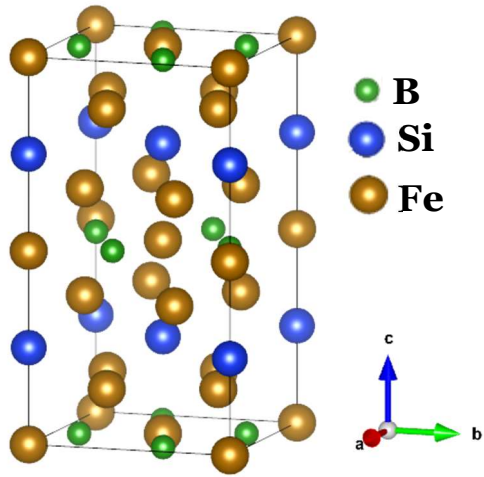

Fig 1: Fe$_5$SiB$_2$ tetragonal crystal structure.

Optimizing intrinsic properties is of fundamental importance for permanent magnets applications: high $T_C$ allows for high working temperatures, while $M_S$ and $H_A$ represent the theoretical limits for remanence and coercivity, respectively. It has been reported, both experimentally and theoretically, that it is possible to tune and increase the anisotropy energy of Fe$_5$SiB$_2$ by chemical substitution either in the metal site by substituting Fe with Co or Mn, or substituting Si in the metalloid site with Ge or P. It has been shown that substituting Si with Ge or P increases anisotropy but leads to the reduction of magnetization and Curie temperature [6], [10], [11]. Similarly, also substituting Fe with Co and Mn suppresses $T_{SR}$ and increases anisotropy but reduces $T_C$ and $M_S$ [12], [13], [14]. It has also been pointed out, by theoretical means, that substituting Fe with Cr could have positive impact on the MAE [15]. Theoretical studies have also suggested that substitution of Re and W in the parent compound Fe$_5$PB$_2$ can increase the anisotropy energy [16], [17], though there are no experimental data available on the effects of Re and Cr chemical substitutions on the Fe$_5$SiB$_2$ system. Ge substituted samples have been prepared and characterized in previous works, but the anisotropy has only been calculated by means of approach to saturation and Sucksmith-Thompson methods [6], [11], [18].

The aim of this work is to synthesize Ge-, Re- and Cr- substituted samples and perform a complete structural and magnetic characterization including direct anisotropy measurements. In particular, we have synthesized and characterized new samples of $Fe_5Si_{1-x}Ge_xB_2$, $Fe_{5-y}Re_ySiB_2$, $Fe_{5-y}Cr_ySiB_2$ with different Ge, Re and Cr contents (x=0, 0.12, 0.25 and y=0, 0.10, 0.25) and we have used macroscopic techniques to measure their effects on the intrinsic magnetic properties of the system, such as Curie Temperature ($T_C$), saturation magnetization ($M_S$) and anisotropy field ($H_A$). We have directly measured the anisotropy field ($H_A$) by means of Singular Point Detection technique (SPD). We have also used local techniques such as nuclear magnetic resonance (NMR) on the boron atom to study the spin reorientation of the material.

## 2. Materials and methods

*2.1 Material synthesis*

$Fe_5Si_{1-x}Ge_xB_2$, $Fe_{5-y}Re_ySiB_2$ and $Fe_{5-y}Cr_ySiB_2$ polycrystalline samples were prepared by arc melting of starting materials Fe (99.9%), Si (99.9%), Ge (99.9%), B (99.9%), Re (99.9%) and Cr (99.9%) under Ar atmosphere in water-cooled copper crucibles. Samples were melted three times and flipped after each melting to improve homogeneity. As-cast button-like samples were annealed under Ar atmosphere at 1173 K for 10 days and quenched in water, they were then annealed again at 1173 K for 7 days and quenched. A total of seven samples with different Ge, Re and Cr contents were prepared. Nominal compositions, sample labels and measured compositions are reported in Table 1.

| Nominal composition | Label | Measured composition |
|---|---|---|
| $Fe_5SiB_2$ | UND | $Fe_{4.96(3)}Si_{1.04(3)}B_2$ |
| $Fe_5Si_{0.88}Ge_{0.12}B_2$ | Ge12 | $Fe_{5.16(2)}Si_{0.78(2)}Ge_{0.06(1)}B_2$ |
| $Fe_5Si_{0.75}Ge_{0.25}B_2$ | Ge25 | $Fe_{5.00(2)}Si_{0.90(2)}Ge_{0.10(1)}B_2$ |
| $Fe_{4.90}Re_{0.10}SiB_2$ | Re10 | $Fe_{4.97(1)}Re_{0.10(1)}Si_{0.93(1)}B_2$ |
| $Fe_{4.75}Re_{0.25}SiB_2$ | Re25 | $Fe_{4.8(2)}Re_{0.23(8)}Si_{0.9(1)}B_2$ |
| $Fe_{4.90}Cr_{0.10}SiB_2$ | Cr10 | $Fe_{5.02(3)}Cr_{0.10(1)}Si_{0.88(3)}B_2$ |
| $Fe_{4.75}Cr_{0.25}SiB_2$ | Cr25 | $Fe_{4.84(5)}Cr_{0.25(3)}Si_{0.91(5)}B_2$ |

Tab 1: Nominal compositions, labels and measured compositions with EDS.

*2.2 Characterization*

Samples were ground into powders for structural characterization by x-ray powder diffractometry with steps of 0.01° in the range from 20° to 90° with a Rigaku Smartlab equipped with copper tube and solid-state sensor hypix3000. GSAS II software was used to analyze the XRD spectra and perform Rietveld refinement. Morphology and chemical composition of polished bulk samples were studied by a Zeiss Auriga Compact scanning electron microscopy (SEM) equipped with an Oxford Xplore 30 energy dispersive x-ray spectroscopy (EDS). EDS compositional analyses were carried out with an acceleration voltage of 20kV. The Curie temperature ($T_C$) was determined from AC susceptibility measurements performed with a homemade AC susceptometer over a temperature range (RT - 1073 K). The applied magnetic field had a frequency of 501 Hz and an amplitude of 10 Oe. Saturation magnetization was measured by a Quantum Design vibrating sample magnetometer (VSM) installed in a physical parameter measurement system (ppms) over the temperature range 4 K - 300 K with a maximum applied field of 5 T. The anisotropy field as a function of temperature was directly measured by Singular Point Detection (SPD) measurements in a pulsed magnetic field (duration t=2.5 ms, max applied field = 6.5 T) over a range of temperatures from 77 K to 300 K [19]. To improve the signal to noise ratio of the SPD signal, powders were dispersed in a thermoplastic adhesive and were oriented in an external constant magnetic field, and finally measured under a pulsed magnetic field applied perpendicularly to the alignment direction, so that the grains' hard directions were parallel to the applied field pulse.

Nuclear magnetic resonance (NMR) measurements were carried out by means of a home-built phase coherent spectrometer and a resonant LC probe head using a nitrogen flow cryostat in zero field [20]. The NMR spectra were obtained point by point by exciting spin echoes at discrete frequencies and the associated spectral amplitudes were determined as the zero-shift Fourier component of the echo signal, divided by the frequency-dependent sensitivity ($\approx \omega^2$). At each frequency step, the LC resonator gets re-tuned by a servo-assisted automatic system plugged into the spectrometer. A standard two pulse $\pi/2$ - $\pi/2$ spin echoes sequence was employed, with rf pulses of intensity and duration optimized for maximum signal, and delay as short as possible, compatibly with the dead time of the apparatus. The pulse length of the NMR experiments was calibrated with the attenuation calibration procedure, since this is the most convenient method in case of broad spectral lines like in the case of magnetic materials [21]. .

### 3. Experimental results

*3.1 Structural characterization and compositional analysis*

X-ray powder diffraction data are displayed in Figure 2 for some representative compositions; a preliminary Rietveld fitting was performed to quantify the effects of composition on the crystal lattice parameters and on the unit cell volume. The crystal structure of the reference sample, a $Cr_5B_3$-type structure, has been used as a template on all the fittings. The simulated patterns are shown as a red line superimposed to the experimental data of Figure 2, and show that the main phase maintains the same crystal structure within all series of samples. The occurrence of a few unindexed peaks in the experimental data highlights the presence of a small quantity of secondary phase, which can be identified as $Fe_5Si_2B$, Fe-B or Fe-Si type depending on the samples. Other secondary phases may be present in quantities too small to be detected by XRD. The complete set of cell parameters for all the samples is reported in Figure 3. The obtained lattice parameters for the Ge substituted sample agree well with the results from [7] and increase gradually with increasing Ge content, this is due to the larger radius of the dopant Ge atoms that replace the smaller Si atoms in the tetragonal crystal structure. Lattice parameters also increase in the new Re substituted compounds while increasing the Cr content reduces both the *a* and *c* lattice parameters, the basal plane being most affected by the substitution. The *c/a* ratio and cell volume follow similar trends, they increase with Re and Ge substitutions and decrease slightly with Cr content (Supplementary material). Compositional analysis was was performed using a scanning electron microscope with an EDS detector. Composition of the main $Fe_5SiB_2$ phase was calculated from the average of ten point measurements on polished bulk samples, boron was neglected in the measurements since it can't be accurately detected because its emission line is very close to the instrumental sensitivity limit. The associated error on composition has been taken as the standard deviation of the ten EDS measurements. Measured compositions together with the sample labels are reported in Table 1.

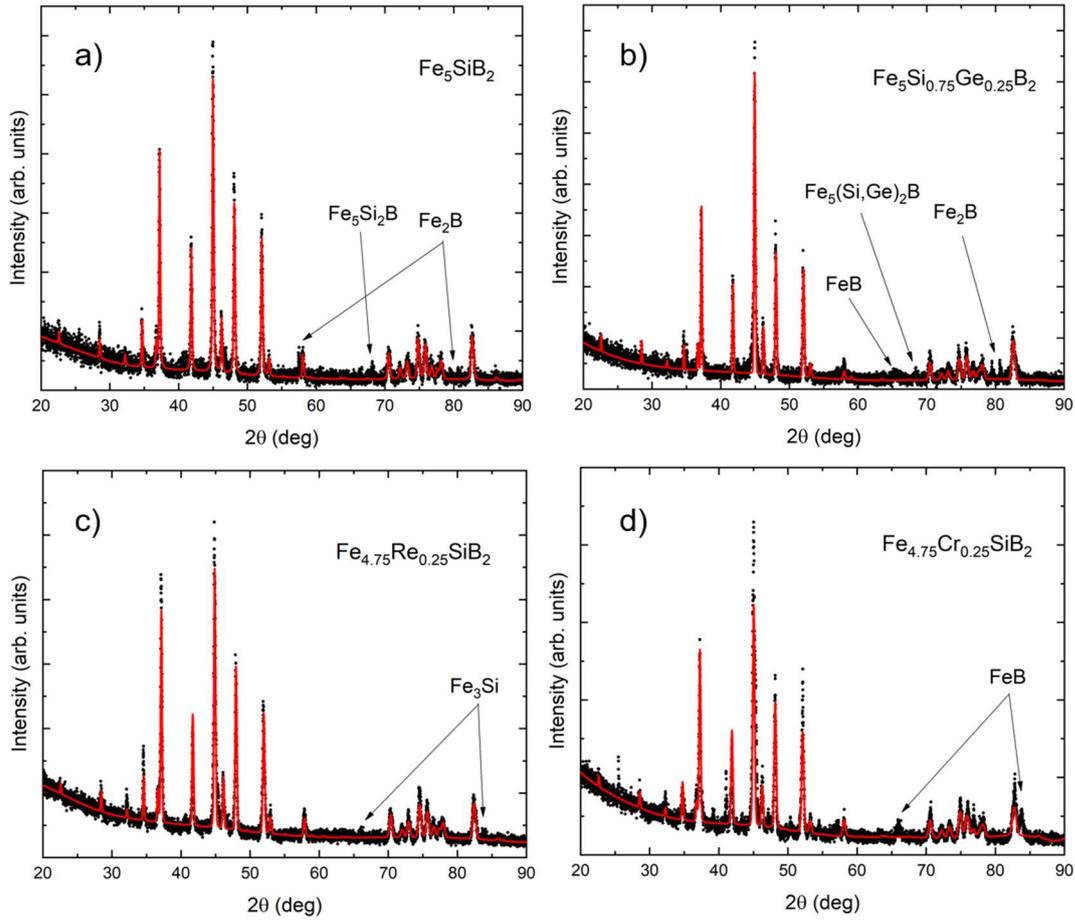

Fig 2: XRD spectra at room temperature for $Fe_5SiB_2$, $Fe_5Si_{0.75}Ge_{0.25}B_2$, $Fe_{4.75}Re_{0.25}SiB_2$ and $Fe_{4.75}Cr_{0.25}SiB_2$.

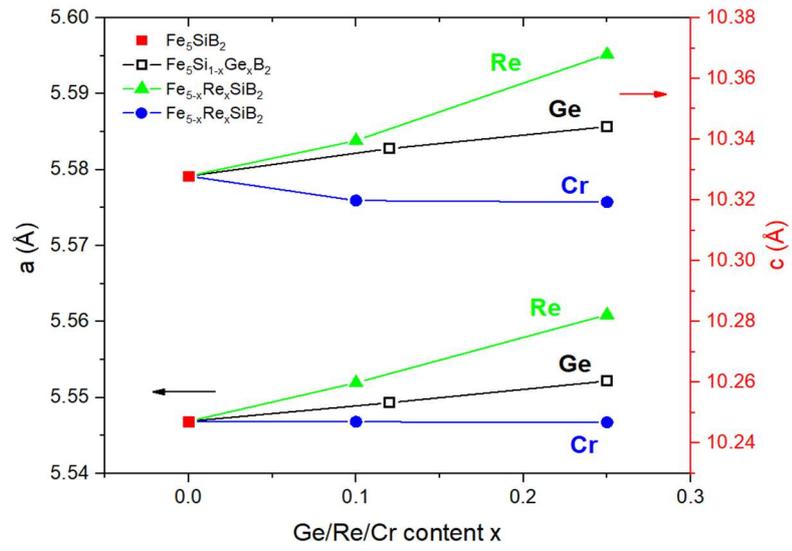

Fig 3: Measured lattice parameters at RT as a function of elemental substitution.

## 3.2 Magnetic characterization

Bulk pieces of samples were cut for magnetic susceptibility measurements as a function of temperature. The Curie temperature ($T_C$) has been determined by the position of the major inflection point in the susceptibility signal χ, corresponding to a minimum in the dχ/dT curve. The susceptibility values as a function of temperature for some representative samples are shown in Figure 4a. The obtained $T_C$ values, reported in Figure 4c, decrease slightly with Ge and Cr content, while Re substitution is more effective in reducing $T_C$ from 803 K down to 753 K. The effect of elemental substitution on magnetization was measured and magnetization curves as a function of applied field for different temperatures are shown in Figure 4b for the Re25 sample and follow similar behavior also in the other samples. The magnetization curves are flat at high field, so we can reasonably approximate the saturation magnetization ($M_S$) value with the magnetization value at 5T. $M_S$ decreases with increasing temperature for all samples, as can be seen in Figure 4e. The $M_S$ values at RT are reported in Figure 4d as a function of elemental substitution. The saturation magnetization at RT of $Fe_5SiB_2$ is 138 $Am^2kg^{-1}$ and is only slightly affected by Ge content, with a reduction of only about 4% at room temperature, while Cr and Re substitutions reduce saturation magnetization by 9% and 14% respectively (Fig 4d). These results on $T_C$ and $M_S$ for Ge substituted samples agree with what has been reported by Clulow et al [6], and the marginal differences can be attributed to differences in phase composition and/or synthesis conditions. Despite all three elemental substitutions reduce $T_C$ and $M_S$, the values obtained are still promising for potential gap magnets. The anisotropy field $H_A$ of the samples was directly measured by means of singular point detection technique (SPD) on oriented powder samples over a temperature range from 77 K to 300 K. According to the theory of SPD for uniaxial systems, in case of easy-axis anisotropy, a peak is present in the $d^2M/dH^2$ curve at a field value corresponding to the anisotropy field. Some representative $d^2M/dH^2$ curves can be found in Supplementary Materials and more details on the SPD technique can be found in [22], [23], [24]. The anisotropy field as a function of temperature presents two different behaviors depending on composition, as can be seen in Figure 4f. In the case of the undoped and Ge doped samples, $H_A$ is constant at low temperature and then increases rapidly when temperature reaches a certain threshold value. In Re and Cr substituted samples, however, $H_A$ continuously increases with temperature. As we will discuss later, the two regimes observed in the undoped and Ge doped samples can be associated with the spin reorientation transition in the material. Ge substitution doesn't have any particular impact on the anisotropy at room temperature, while Re and Cr increase $H_A$ by about 15% and 6% respectively.

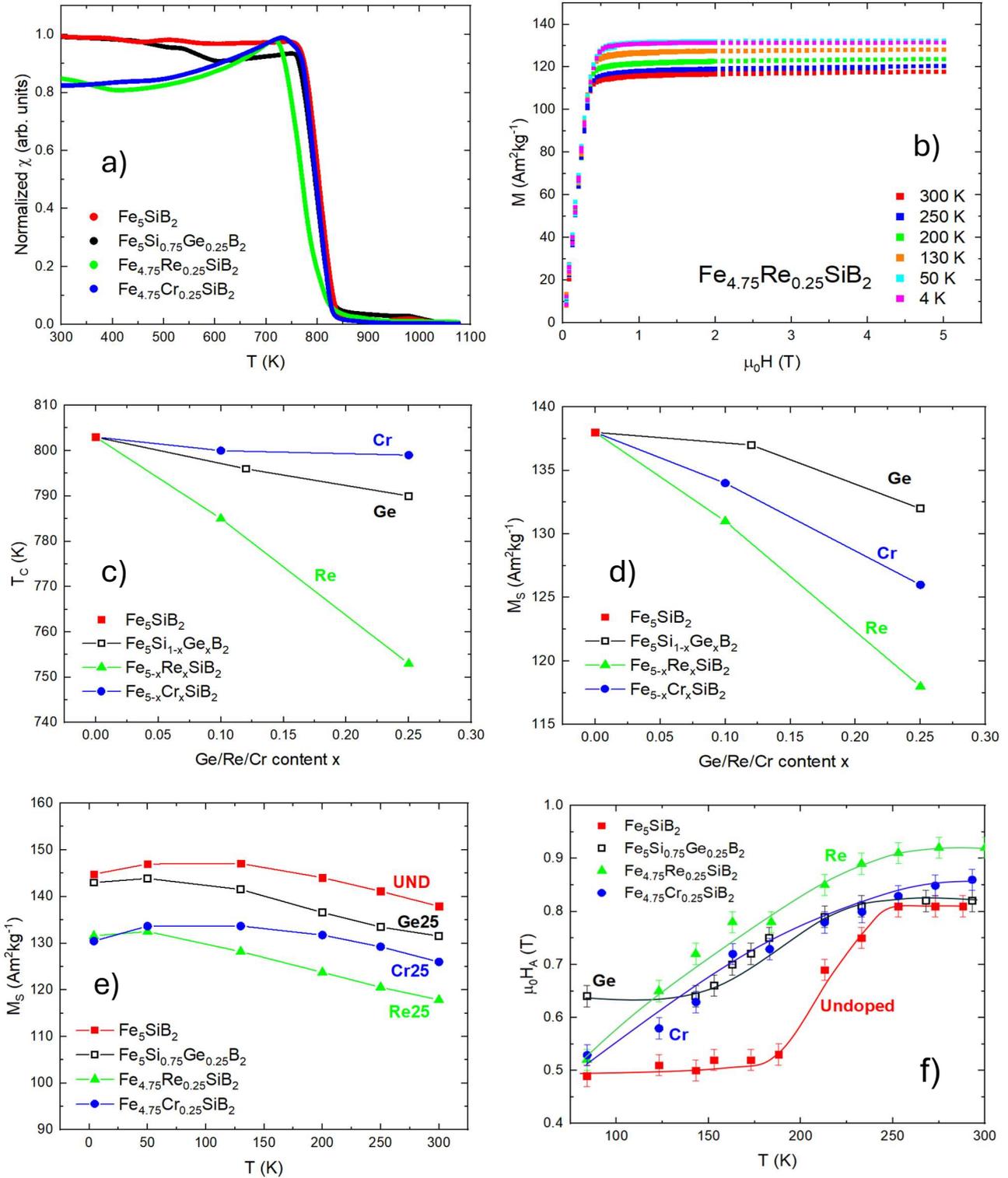

Fig 4: (a) Magnetic susceptibility χ as a function of temperature. (b) Magnetization as a function of applied field at different temperatures. (c) Curie temperature, $T_C$, as a function of elemental substitution. (d) Saturation magnetization, $M_S$, as a function of elemental substitution and (e) as a function of temperature. (f) Anisotropy field, $H_A$, as a function of temperature as measured by SPD, lines are guides for the eye.

*3.3 Nuclear Magnetic Resonance measurements*

Hyperfine-field NMR measurements in zero applied field were carried out on powders of the undoped sample and of the three samples with maximum content of Ge, Re and Cr from 77 K to room temperature. Broad zero-field NMR spectra were recorded in the 22-35 MHz frequency interval (Figure 5). The occurrence of a spectral replica of these signals close to 10 MHz, nicely scaled in frequency by a factor of 2.986 corresponding to the ratio of the nuclear g-factors of the $^{11}B$ and $^{10}B$ boron isotopes (Supplementary material), unambiguously assigns them to the resonance of $^{11}B$ ($\gamma_{11}/2\pi$ = 13.66 MHz/T) in a spontaneous hyperfine field of ≈2T at the boron nucleus. Notably, in the same frequency range also $^{57}Fe$ NMR peaks are predicted from Mössbauer spectroscopy [8]. However, their accidental overlap with the $^{11}B$ NMR spectrum makes them undetectable, due to the much larger receptivity of $^{11}B$ relative to $^{57}Fe$. Optimal excitation of the $^{11}B$ NMR signals was achieved with very low rf level relative to the power required for bare nuclei, indicative of the presence of a strong rf enhancement. The latter, namely, the amplification factor between the effective oscillating field $H_1^*$ at the nucleus and the applied rf field $H_1$, due to the electronic magnetic susceptibility of the system and the hyperfine coupling of the nuclear spins, was estimated in the order of $\eta = H_1^*/H_1 \approx 1000$, a value typical for nuclei in domain walls of an unsaturated ferromagnet [21].

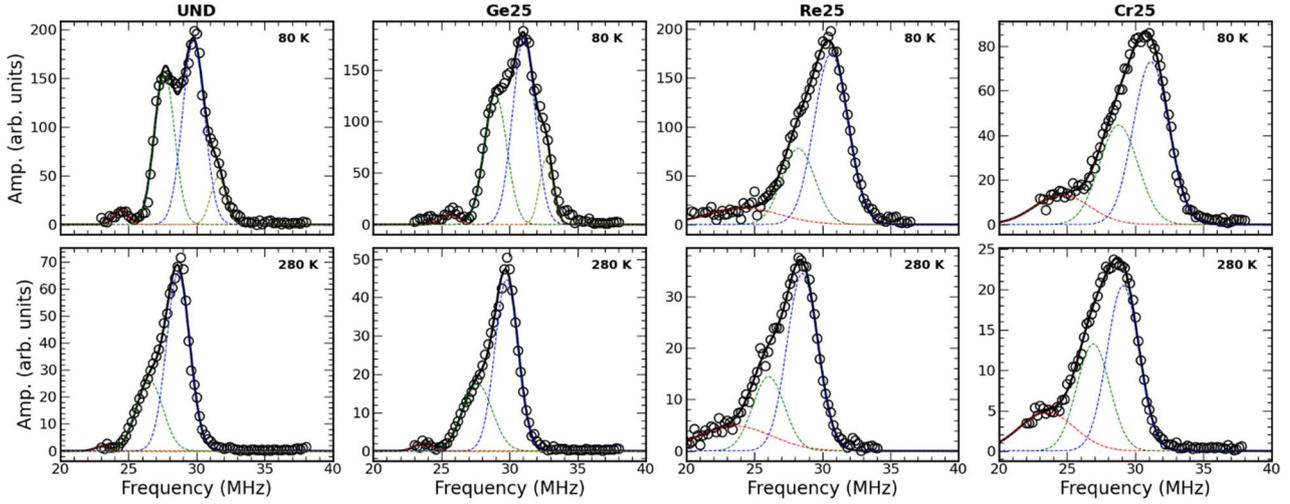

Fig 5: NMR spectra for different elemental substitutions undoped (a), Ge (b), Re (c), Cr (d), obtained at different temperatures.

The spectra of the different compositions at room temperature and 80 K are reported in Figure 5. They can be reproduced phenomenologically by a multi-Gaussian function:

$$f(x) = \sum_{i=1}^{N} \frac{A_i}{\sigma_i \sqrt{2\pi}} e^{-\frac{(x-v_i)^2}{2\sigma_i}} \quad (2)$$

where $N=3$ unresolved components account accurately for the asymmetric shape observed for all the samples at room temperature and, in the Re and Cr-doped ones, also at lower temperatures. The spectra of the undoped and Ge-doped samples at low temperature, however, exhibit a clear pattern of peaks requiring a fit with $N=4$ components. The complex spectral shapes, which contrast with the single boron site of the crystal structure, can be understood qualitatively on the basis of an anisotropic hyperfine coupling of B nuclei and the rotation of the electronic spins inside domain walls, whence the NMR signal originates. It can be shown, in particular, that the smooth asymmetric shapes observed at room temperature can be reproduced by an isotropic spin rotation in the presence of a cylindrical hypefine tensor (Supplementary material). The peculiar shapes of the low-temperature spectra in undoped and Ge-doped $Fe_5B_2Si$ denote the transition to a different micromagnetic structure in the unsaturated material, whereby spins inside a domain wall do not span isotropically the solid angle. Such a transition is apparently related to the change in the magnetic anisotropy and the spin reorientation detected by magnetometry [9].

The temperature evolution of the spin reorientation occurring in the undoped and Ge-doped materials above 80 K, at variance with the Cr and Re-doped ones, can be traced phenomenologically by the temperature dependence of the first and third moment of the spectra. The average resonance frequency $v(T)$ of the undoped and Ge-doped samples exhibits a clear anomaly at approximately 140K, in place of the smooth decrease with increasing temperature as a regular order parameter observed in the other samples (Figure 6a). The reorientation is also apparent as a variation in the skewness of the spectra, smoothed over a temperature interval (Figure 6b). In the undoped and Ge-doped samples the third moment obeys a law

$$M_3(T) = a + \frac{b}{2}\left(1 + \mathrm{erf}\left(\frac{T-T_{SR}}{\sigma}\right)\right) \quad (3)$$

with $T_{SR}$ and σ the mean transition temperature and its distribution width, respectively. Their best fit parameters, tabulated in Table 2, are in fair agreement with the SPD data. It is also noteworthy a qualitative change in the shape of the spectra at $T_{SR}$, with a clearer peak pattern at lower temperature, whose accurate fit requires an extra component (Figure 5 and Supplementary material).

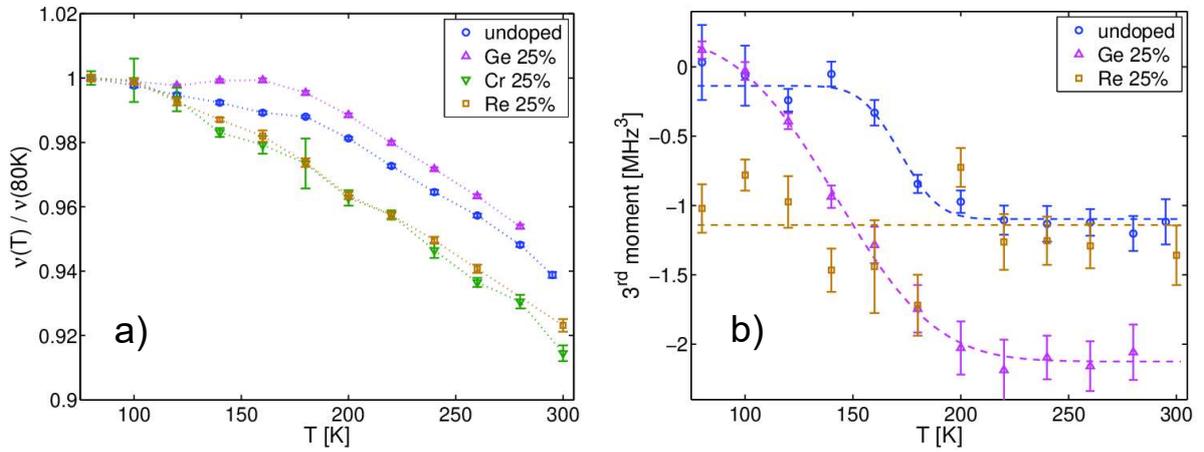

Fig. 6: (a) First moment of the $^{11}$B NMR spectra of the four samples, normalized vs. the 80K value, as a function of temperature. (b): Third moments of the spectra vs. temperature, for the undoped, Ge- and Re-doped compounds. Data for the Cr-doped are similar to those of the Re-doped one and are omitted for clarity.

## 4. Discussion

XRD and EDS measurements proved that Ge, Re and Cr entered in the tetragonal lattice, as shown in Table 1 and Figure 3. The presence of small quantities of secondary phases (FeB and Fe$_2$B) might have some marginal effect in increasing the saturation magnetization values, but the general trend due to the effect of dopant elements is unchanged. Anisotropy values are not affected by this because the secondary phase content is too low to be detected by the SPD experimental set-up. We note that Ge and Cr substitutions do not have any particular effect on the Curie temperature of the compound, and they decrease T$_C$ by less than 2%. Substitution of Re has a stronger effect on T$_C$, which is reduced by about 6% down to 753 K (Figure 4c). Re also has the strongest effect in reducing the saturation magnetization, while Ge and Cr have minor effects. However, despite this reduction, both the Curie temperature and the saturation magnetization maintain significant values that are desirable for potential gap magnets (Figure 4d). As shown in Figure 4f, the anisotropy field at room temperature is increased up to 0.86 T and 0.92 T by the substitution of Cr and Re respectively, but it is not improved by Ge substitution. In the case of Ge containing samples, these observations are in line with previous results in literature regarding $T_C$ and $M_S$, but we don't detect any positive effect of Ge on the room temperature anisotropy as instead reported in [6], [11].

Another important dimensionless intrinsic parameter is the magnetic hardness parameter κ, defined as:

$$\kappa = \sqrt{\frac{K_1}{\mu_0 M_S^2}} \tag{1}$$

where $K_1 = (H_A \cdot M_S)/2$ is the magnetocrystalline anisotropy constant, $\mu_0$ is the vacuum magnetic permeability and $M_S$ is the saturation magnetization. If $\kappa > 1$, the material can sustain self-demagnetization and magnets of any shape can be produced [3]. If $\kappa < 1$, applications are possible only for magnets with specific shapes that maximize coercivity such as elongated spheroids and needles. The $Fe_5SiB_2$ system as well as the Ge, Re and Cr substituted compounds have κ in the range $0.58 < \kappa < 0.67$, thus they can be classified as semi-hard materials, and their applications would be limited by shape.

*4.1 Spin reorientation transition*

The anisotropy field as a function of temperature presents two different behaviors. For the Re doped and Cr doped samples it increases continuously with temperature, while for the undoped and Ge doped samples it is constant at low temperature and then rapidly increases at higher temperatures (Figure 4f). The presence of two regimes can be attributed to the spin reorientation transition previously observed at 172 K in the undoped parent compound [11]. The spin reorientation transition is due to the temperature dependence of the anisotropy constants, in particular $K_1$. At low temperature, $K_1$ is negative and the material has an easy plane of magnetization. When the temperature reaches the spin reorientation temperature $T_{SR}$, $K_1$ becomes positive, the *c* axis becomes the easy axis of magnetization and the value of anisotropy increases rapidly. The temperature range where an increased anisotropy is observed corresponds to the spin reorientation temperature of the material. From the anisotropy measurements, we can say that the undoped and Ge doped samples both show a spin reorientation at about 190 K and 140 K respectively. These results are also in agreement with observations from literature that report the partial suppression of $T_{SR}$ upon Ge substitution [6], [11]. On the other hand, in Re and Cr doped compounds, the anisotropy field continuously increases with temperature because $K_1$ only changes in magnitude but never changes in sign, thus the spin reorientation transition is suppressed. $T_{SR}$ can also be measured from the magnetization curves at low field as shown in Supplementary materials. When $K_1$ goes from negative to positive, it crosses the zero value. For small values of $K_1$ the material has a greater response to the applied field. This produces a peak in the magnetization curve close to the transition temperature.

The spin reorientation transition has also been highlighted by $^{11}B$ NMR, whose zero-field spectra probe the spin distribution inside domain wall via the anisotropic term of the hyperfine coupling of boron. The shape of the spectra, as well as spectral indicators such their odd moments, coherently indicate, in the undoped and Ge doped compounds, a qualitative change in the structure of the domain walls occurring between 100 and 200 K, reflecting a change in the magnetic anisotropy driving the spin reorientation. However, such a transition is not detected above 77 K in the Re and Cr doped samples, whose spectra are qualitatively identical at the lowest and highest temperature, as shown in Figure 5. This observation is consistent with the suppression of spin reorientation in those compounds observed by SPD. A similar spin reorientation suppression has also been reported for $Fe_5PB_2$ and Mn and Co substituted $Fe_5SiB_2$ [13]. A reliable estimate of the spin reorientation transition is apparently provided by the third moment of the NMR spectra, yielding mean $T_{SR}$ values of 172 K and 142 K in the undoped and Ge doped compounds, respectively, which are consistent with SPD results within the experimental errors (Table 2). In particular, both methods confirm that Ge doping shifts the spin reorientation to lower temperatures.

| Sample | Method | $T_{SR}$ (K) | σ (K) |
|---|---|---|---|
| $Fe_5SiB_2$ | SPD | 190(10) | |
| | NMR | 172(3) | 20(5) |
| $Fe_5Si_{0.75}Ge_{0.25}B_2$ | SPD | 140(10) | |
| | NMR | 142(2) | 50(4) |

Tab 2: Spin reorientation temperatures for the undoped and Ge doped samples. The width σ of the transition from NMR data is also reported. Re and Cr-doped samples don't show any spin reorientation.

## 5. Conclusions

We have performed a thorough study of the structural and intrinsic magnetic properties of Ge, Re and Cr substituted $Fe_5SiB_2$. Noticeably, the magnetic anisotropy was directly measured by application of SPD technique in pulsed magnetic field. Results show that the samples crystallize in the tetragonal $Cr_5B_3$-type structure with the I4/mcm space group. Ge substitution only has a minor impact on Curie temperature and saturation magnetization, while the anisotropy at room temperature is unchanged. Cr substitution reduces slightly the saturation magnetization and improves the room temperature anisotropy by about 6%. The substitution of Re is the most effective in reducing both $T_C$ and $M_S$ but also increases the room temperature anisotropy by about 15%. The spin reorientation temperature was measured by means of two different methods using SPD and NMR, in agreement also with magnetization curves at low field. Re and Cr substitutions completely suppress the transition while Ge only partially reduces $T_{SR}$ to about 140 K. Despite the reduction of $T_C$ and $M_S$ upon elemental substitution, the obtained values are still promising for potential gap magnets. However, the obtained anisotropy values are not enough to achieve a magnetic hardness parameter κ > 1. This is not ideal for practical applications because it doesn't fulfill the general criterion on intrinsic magnetic properties for obtaining permanent magnets.

Possible applications of this material would be limited only to certain shapes with low demagnetizing factors, such as elongated spheroids or needles [5].

## Acknowledgements

This work was financed by the European Union - NextGenerationEU (National Sustainable Mobility Center CN00000023, Italian Ministry of University and Research Decree no. 1033, 17/06/2022, Spoke 11, Innovative Materials & Lightweighting). The opinions expressed are those of the authors only and should not be considered as representative of the European Union or the European Commission's official position. Neither the European Union nor the European Commission can be held responsible for them.

# Bibliography

[1] H. Nakamura, «The current and future status of rare earth permanent magnets», *Scr. Mater. 154 273-276*, 2018.

[2] S. Sugimoto, «Current status and recent topics of rare-earth permanent magnets», *J. Phys. Appl. Phys. 446*, 2011.

[3] J. M. D. Coey, *Magnetism and magnetic materials*, Repr. Cambridge: Cambridge Univ. Press, 2013.

[4] J. M. D. Coey, «Perspective and Prospects for Rare Earth Permanent Magnets», *Eng. 62 119-131*, 2020.

[5] J. M. D. Coey, «Permanent magnets: Plugging the gap», *Scr. Mater. 676 524-529*, 2012.

[6] R. Clulow, «Magnetic and structural properties of the $Fe_5Si_{1-x}Ge_xB_2$ system», *J. Solid State Chem. 316*, 2022.

[7] «Aronsson, B. E. R. T. I. L., et al. "X-ray Investigations on Me-Si-B Systems (Me= Mn, Fe, Co)." Acta Chem. Scand 14.6, 1960».

[8] R. Wäppling, T. Ericsson, L. Häggström, e Y. Andersson, «Magnetic Properties of $Fe_5SiB_2$ and related compounds», *J. Phys. Colloq. 37 C6-591*, 1976.

[9] J. Cedervall, «Magnetostructural transition in $Fe_5SiB_2$ observed with neutron diffraction», *J. Solid State Chem. 235 113-118*, 2016.

[10] D. Hedlund *et al.*, «Magnetic properties of the $Fe_5SiB_2$–$Fe_5PB_2$ system», *Phys. Rev. B 969*, 2017.

[11] B. T. Lejeune, «Synthesis and processing effects on magnetic properties in the $Fe_5SiB_2$ system», *J. Alloys Compd. 731 995-1000*, 2018.

[12] J. Cedervall *et al.*, «Influence of Cobalt Substitution on the Magnetic Properties of $Fe_5PB_2$», *Inorg. Chem. 572 777-784*, 2018.

[13] M. A. McGuire e D. S. Parker, «Magnetic and structural properties of ferromagnetic $Fe_5PB_2$ and $Fe_5SiB_2$ and effects of Co and Mn substitutions», *J. Appl. Phys. 11816*, 2015.

[14] M. Werwinski *et al.*, «Magnetic properties of $Fe_5SiB_2$ and its alloys with P, S, and Co», *Phys. Rev. B 9317*, 2016.

[15] J. Thakur, P. Rani, e M. Tomar, «Tailoring in-plane magnetocrystalline anisotropy of $Fe_5SiB_2$ with Cr-substitution», *AIP Conf. Proc. Vol 2115 No 1 AIP Publ.*, 2019.

[16] J. Thakur, P. Rani, e M. Tomar, «Enhancement of magnetic anisotropy of $Fe_5PB_2$ with W substitution: ab-initio study», *AIP Conf. Proc. Vol 2093 No 1 AIP Publ.*, 2019.

[17] M. Werwinski *et al.*, «Magnetocrystalline anisotropy of $Fe_5PB_2$ and its alloys with Co and 5d elements: A combined first-principles and experimental study», *Phys. Rev. B 9821*, 2018.

[18] R. Hirian, «Investigations on the magnetic properties of the $Fe_{5-x}Co_xSiB_2$ alloys by experimental and band structure calculation methods», *Ournal Magn. Magn. Mater. 505*, 2020.

[19] G. Asti e S. Rinaldi, «Singular points in the magnetization curve of a polycrystalline ferromagnet», *J. Appl. Phys. 458*, 1974.

[20] G. Allodi, A. Banderini, R. De Renzi, e C. Vignali, «HyReSpect: A broadband fast-averaging spectrometer for nuclear magnetic resonance of magnetic materials», *Rev. Sci. Instrum. 768*, 2005.

[21] C. Meny e P. Panissod, «Nuclear magnetic resonance in ferromagnets: Ferromagnetic nuclear resonance; a very broadband approach», in *Annual Reports on NMR Spectroscopy. 103. Academic Press, 47-96*, Elsevier, 2021.

[22] G. Asti, F. Bolzoni, e R. Cabassi, «Singular point detection in multidomain samples», *J. Appl. Phys. 731 323-333*, 1993.

[23] F. Bolzoni e R. Cabassi, «Review of singular point detection techniques», *Phys. B Condens. Matter 346 524-527*, 2004.

[24] R. Cabassi, «Singular Point Detection for characterization of polycrystalline permanent magnets», *Meas. 160*, 2020.

Supplementary Material for

# Magnetic properties of Ge, Re and Cr substituted $Fe_5SiB_2$

M. Casadei[a,b*], M.M. Isah[a], R. Cabassi[b], G. Trevisi[b], S. Fabbrici[b], C.D. Fernández[b], G. Allodi[c], V. Fournée[d], S. Sanna[a], F. Albertini[b]

[a] Department of Physics, University of Bologna, viale Berti Pichat 6/2, Bologna, Italy
[b] IMEM-CNR, Via delle Scienze 37, Parma, Italy
[c] Department of Mathematical, Physical and Computer Science, University of Parma, Parco Area delle Scienze 7/A, Parma, Italy
[d] Institut Jean Lamour, Campus Artem, 2 allée André Guinier, 54011 Nancy, France

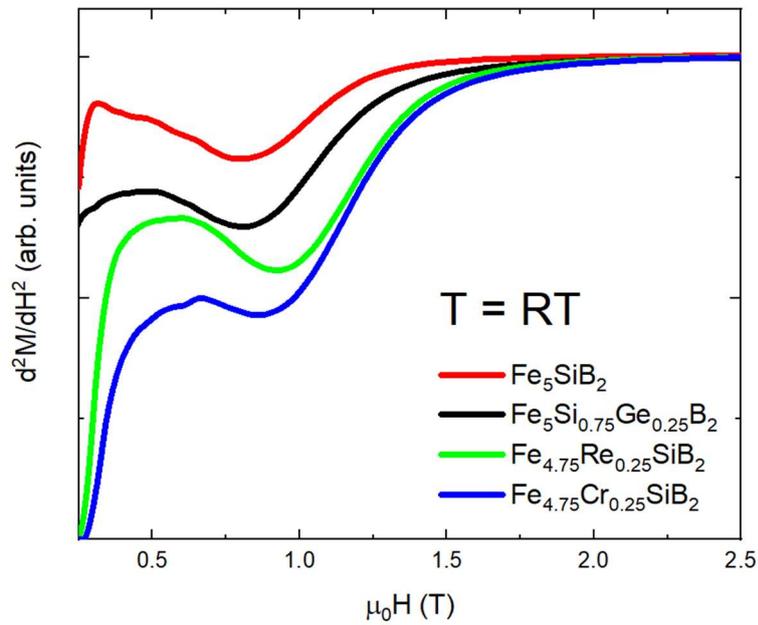

Fig S1: SPD curves of $d^2M/dH^2$ as a function of applied field for $Fe_5SiB_2$ and Ge, Re and Cr substituted samples at room temperature. The position of peak corresponds to the anisotropy field of the material.

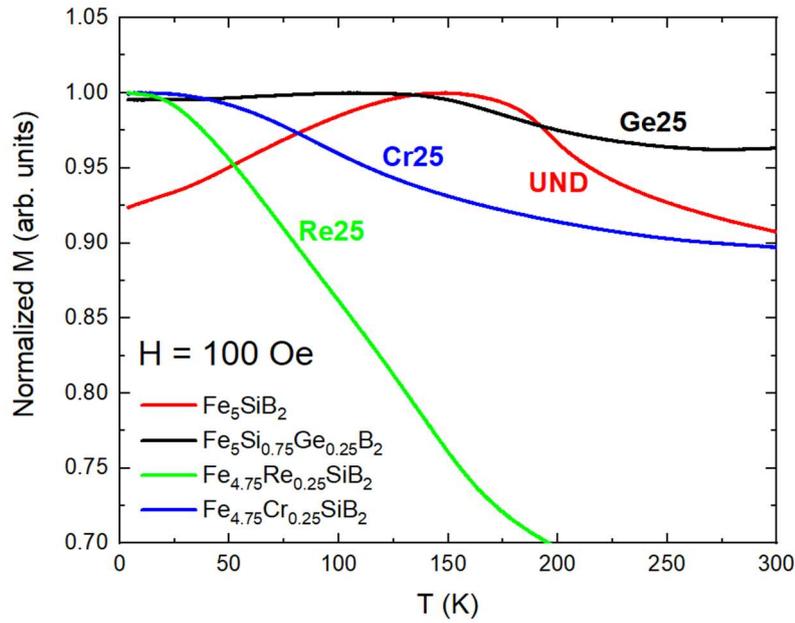

Fig S2: Low field (100 Oe) magnetization curves for $Fe_5SiB_2$ and Ge, Re and Cr substituted samples as a function of temperature. Notice how the undoped and Ge doped samples show peaks at different temperatures, corresponding to the spin reorientation temperature, while Re and Cr doped samples have a monotonous behavior.

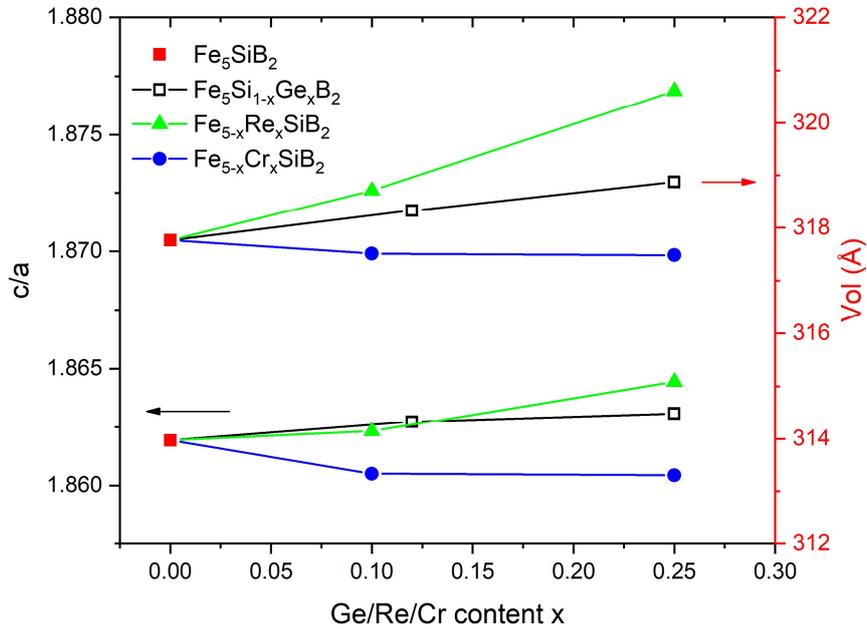

Fig S3: c/a ratio and cell volume as a function of elemental substitution.

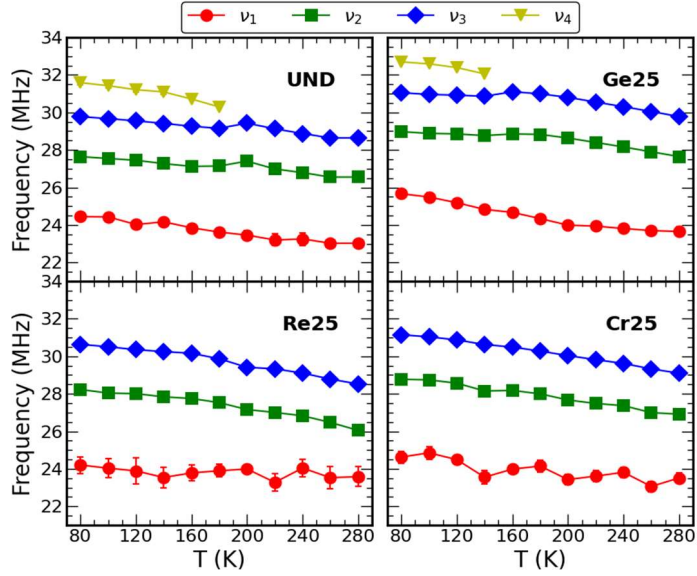

Fig S4: Frequencies of the Gaussian components needed to fit the spectra at different temperatures for a) undoped, b) Ge doped, c) Re doped and d) Cr doped samples with the maximum content of substituting elements.

## Evidence for $^{11}$B NMR

The experimental proof that the reported MNR signals are due to the $^{11}$B resonance is provided by the occurrence of replica spectra for the $^{10}$B isotope. As both isotopes experience the same hyperfine field, their resonance frequency are scaled by their gyromagnetic ratio, $\gamma_{11}/2\pi$ = 13.66 MHz/T and $\gamma_{10}/2\pi$ = 4.5743 MHz/T for $^{11}$B and $^{10}$B, respectively.

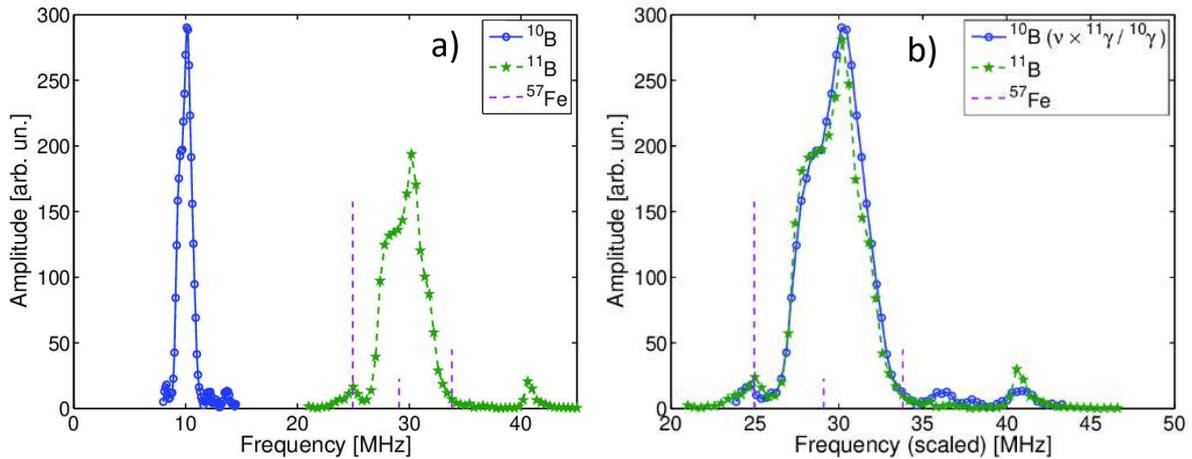

Figure S5: a) Zero-field NMR frequency spectrum of $^{10}$B (bullets) and $^{11}$B stars of undoped Fe$_5$SiB$_2$ at 77 K. b) The two spectra overlaid, with the $^{11}$B spectrum unscaled, and frequencies of the $^{10}$B spectral points multiplied by $\gamma_{11}/\gamma_{10}$ = 2.986, where $\gamma_{10}$, $\gamma_{11}$ are the gyromagnetic ratios of the $^{10}$B and $^{11}$B boron isotopes, respectively. In both panels, the dashed vertical lines mark the $^{57}$Fe resonance frequencies predicted from the $^{57}$Fe Moessbauer spectra of Wäppling et al. [7].

## Model for the NMR spectral shape

Here we outline a simple model for the shapes of the $^{11}$B NMR spectra. The model assumes an anisotropic hyperfine interaction

$$\nu = \sqrt{(A_x^2 \cos^2 \phi + A_x^2 \sin^2 \phi) \sin^2 \theta + A_x^2 \cos^2 \theta}$$

where θ, φ are the polar angles of the spin relative to the crystal axes, and a random orientation of the spin inside a domain wall, whereby the solid angle is spanned with uniform probability. The spectral weight $g(\nu)$ thus coincides with the solid angle portion corresponding to a resonance frequency ν. In the presence of a cylindrical hyperfine tensor ($A_x = A_y$) the spectral density can be calculated analytically as

$$g(\nu) = \frac{1}{\left|\frac{d\nu}{d\cos\theta}\right|} = \frac{\nu}{\sqrt{(\nu^2 - A_x^2)(A_z^2 - A_x^2)}}$$

showing a powder singularity at $\nu = A_x$. The latter, however, is smeared out in practice by any source of inhomogeneity in the hyperfine coupling. In the general case, a numerical integration is needed to calculate $g(\nu)$ and a powder singularity does not show up for non degenerate hyperfine tensor components $A_x \neq A_y \neq A_z$.

In Fig. S6 we show a fit of the spectrum of the undoped compound well above the spin reorientation ($T = 280$K) to the model sketched above, where the inhomogeneity of the hyperfine coupling is accounted for by an incoherent Gaussian broadening (i.e. the convolution of the calculated spectrum with a Gaussian distribution). The fit reproduces the skew lineshape of the spectrum with reasonable accuracy, yielding best fit parameters $A_x = A_y = 29.27(5)$ MHz, $A_z = 25.19(2)$ MHz and Gaussian broadening σ = 0.93(3) MHz at this particular temperature in this sample. Fits of similar quality are obtained for the Ge sample above the reorientation transition and for the other samples not showing it.

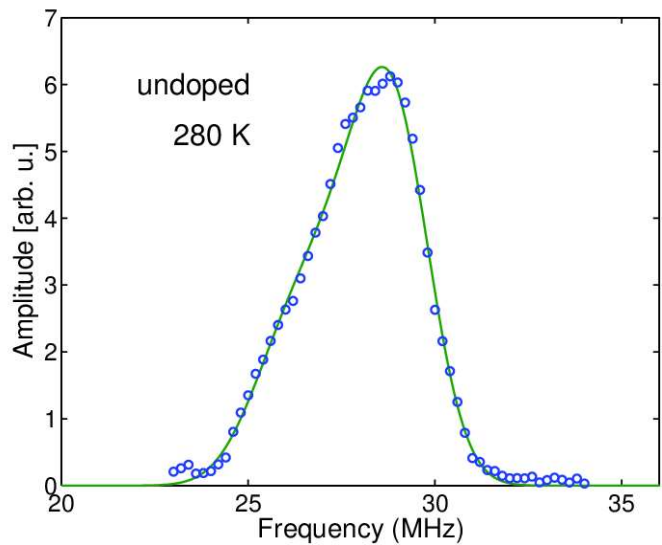

Figure S6: Zero-field NMR spectrum of the undoped sample at 280K, and its best fit to an isotropic angular spin spin distribution in the presence of an anistropic hyperfine coupling, as discussed in the text.